\documentclass[pra,twocolumn,showpacs,superscriptaddress,byrevtex]{revtex4}
\usepackage{amsfonts}
\usepackage{amsmath}
\usepackage{amssymb}
\usepackage[dvips]{graphicx}
    \newcommand{\ket}[1]{\ensuremath{|\,{#1}\,\rangle}}
    \newcommand{\bra}[1]{\ensuremath{\langle\,{#1}\,|}}
    \newcommand{\braket}[2]{\ensuremath{\langle\,{#1}\,|\,{#2}\,\rangle}}

    \newcommand{\lsub}[1]{\ensuremath{_{_{\!\scriptstyle #1}}}}

    \newcommand{\itg}[1]{\ensuremath{\int\!\!d{#1}\!\!}}
    \newcommand{\itgf}[1]{\ensuremath{\int\!\!d{#1}\,}}

    \newcommand{\sinc}{\ensuremath{\mbox{\hspace{1.3pt}sinc}\,}}

    \newcommand{\re}[1]{\ensuremath{\mbox{\hspace{1.3pt}Re}}({#1})}
    \newcommand{\im}[1]{\ensuremath{\mbox{\hspace{1.3pt}Im}}({#1})}

\begin{document}

\title{State reconstruction for composite systems of two spatial qubits}

\author{G. Lima\footnote{Present address: Group of Applied Physics, University of Geneva, 1211 Geneva 4,
Switzerland.}}
\address{Departamento de F\'{\i}sica,
Universidade Federal de Minas Gerais, Caixa Postal 702,
Belo~Horizonte, MG 30123-970, Brazil.}
\email{glima@cfm.cl}
\author{F.A. Torres-Ruiz}
\address{Center for Quantum Optics and Quantum Information,
Departamento de F\'{\i}sica, Universidad de Concepci\'on,  Casilla
160-C, Concepci\'on, Chile.}
\author{Leonardo Neves
\footnote{Present address: Clarendon Laboratory, University of
Oxford, Parks Road Oxford Ox 13 PU, UK.}}
\address{Departamento de F\'{\i}sica,
Universidade Federal de Minas Gerais, Caixa Postal 702,
Belo~Horizonte, MG 30123-970, Brazil.}
\author{A. Delgado}
\address{Center for Quantum Optics and Quantum Information,
Departamento de F\'{\i}sica, Universidad de Concepci\'on,  Casilla
160-C, Concepci\'on, Chile.}
\author{C. Saavedra}
\address{Center for Quantum Optics and Quantum Information,
Departamento de F\'{\i}sica, Universidad de Concepci\'on,  Casilla
160-C, Concepci\'on, Chile.}
\author{S. P\'adua}
\address{Departamento de F\'{\i}sica,
Universidade Federal de Minas Gerais, Caixa Postal 702,
Belo~Horizonte, MG 30123-970, Brazil.}

\begin{abstract}
Pure entangled states of two spatial qudits have been produced by
using the momentum transverse correlation of the parametric
down-converted photons [Phys. Rev. Lett. \textbf{94} 100501]. Here
we show a generalization of this process to enable the creation of
mixed states of spatial qudits and by using the technique proposed
we generate mixed states of spatial qubits. We also report how the
process of quantum tomography is experimentally implemented to
characterize these states. This tomographic reconstruction is
based on the free evolution of spatial qubits, coincidence
detection and a filtering process. The reconstruction method can
be generalized for the case of two spatial qudits.
\end{abstract}

\pacs{03.65.-w, 03.67.Mn, 42.65.Lm, 03.65.Wj, 03.67.Hk}


\maketitle

\section{Introduction}
\label{sec:intro}

The concept of quantum state plays a central role in the Quantum
Theory. It is considered to be the most complete description for a
physical system. Statistical distributions of results of
experiments carried out on a physical system can be completely
predicted from its initial state. Therefore, the experimental
determination of an initially unknown quantum state becomes a very
important research subject. This has led to the development
of techniques to perform  the state determination. In recent years
the problem of state determination has received a considerable
degree of attention due to important results in Quantum
Information Theory.

Several techniques have been designed and used for the state
estimation of different physical systems. In the field of atomic
physics, quantum endoscopy was used to determine the state of ions
and atoms \cite{Vogel1,Schleich,Walmsley}. In quantum optics, the
Wigner function of multi mode fields could be measured using
homodyne detection \cite{Vogel2,Smithey,Vogel3} and the technique
of quantum tomographic reconstruction (QTR) was used for measuring
the polarization state of parametric down-converted photons
\cite{White}.

In general, these methods are based on a linear inversion of the
measured data. In the case of QTR, the data is acquired with a
series of measurements performed on a large number of identically
prepared copies of a quantum system. The fact that this
transformation is linear, makes it strongly dependent of any
experimental error that may occur while recording the data. It can
appear as a consequence of the experimental noise or misalignment
and therefore, the reconstructed state is only a reasonable
approximation of the real quantum state. The density matrices
obtained may have properties that are not fully compatible with a
quantum state. Another alternative that has been considered for
the state determination is the numerical technique called maximum
likelihood estimation \cite{Hradil,James}. It is based on a
relation between the measured data and the quantum state that
could have generated them. Even though it generates only possible
density matrices, it has the drawback of enhancing the uncertainty
on the state estimation.

In this article we are interested in the determination of the
state of a composite system. In our experiment, the system
corresponds to two photons generated by spontaneous parametric
down-conversion (SPDC). In this nonlinear process a photon from a
pump laser beam incident to a non-linear crystal originates
probabilistically two photons, signal and idler \cite{MandelBook}.
The photons of this pair are also called twin photons for being
generated simultaneously \cite{HOM-interferometer}. Recently, we
have demonstrated that by placing $D$ symmetric slits at the path
of each twin photon it is possible to use the transverse
correlations of the photon pair to generate maximally entangled
states of two effective $D$-dimensional quantum systems
\cite{Leonardo,GLima,otherqudits}. We refer to these
$D$-dimensional quantum systems as spatial qudits. In the present
work we extend previous results to the generation of mixed states
of two spatial qubits, which also applies to the case of qudits.
Following this, we investigate the state determination of two
spatial qubits. We show both theoretically as well as
experimentally, that one can implement the process of QTR to
obtain the density operator of a state composed of two spatial
qubits. The quality of the reconstruction performed is also
discussed. Even though we had considered only the special case of
spatial qubits, it is straightforward to show that the technique
used can be generalized for being applied to a system composed of
two spatial qudits. The main motivation on studying both the
generation and the reconstruction of mixed states of spatial
qudits is to consider more realistic experimental situations in
case of using them in technological fields, such as quantum
communication, where pure states can become into mixed ones due to
interactions with their environment.

\section{Controlled generation of mixed states}
\label{sec:generation}

It was shown in \cite{Leonardo,GLima} that the state of parametric
down-converted photons when each photon is transmitted through
identical multi slits is given by
\begin{equation}
\label{stgeral} \ket{\Psi} = \sum_{l=-l_D}^{l_D} \; \sum_{m=-l_D}^{l_D} W_{lm} \;
\exp\left(i\frac{kd^{2}}{8z_{A}}(m-l)^2\right) \; \ket{l}\lsub{s} \otimes
\ket{m}\lsub{i},
\end{equation}
where $D$ is the number of slits in each multi slits array, $d$ is
the distance between the center of two consecutive slits, $a$ is
the half width of the slits and $l_D = (D-1)/2$. The function
$W_{lm}$ is the spatial distribution of the pump beam at the plane
of the multi slits ($z=z_{A}$) and at the transverse position
$x=(l+m)d/2$,
\begin{equation}
\label{wlm}
W_{lm} = W \left[\frac{(l+m)d}{2};z_A\right].
\end{equation}

The $\ket{l}_{s}$ (or $\ket{m}_{i}$) state is a single-photon state defined, up to a
global phase factor, by the expression
\begin{equation}
\label{base}
\ket{l}\lsub{j} \equiv \sqrt{\frac{a}{\pi}} \itgf{q_{j}} \exp(-iq_{j}ld) \sinc(q_{j}a)\ket{1q_{j}},
\end{equation}
and represents the photon in mode $j$ ($j = i, s$) transmitted by
the slit $l$. The transverse component of wave vector of the
down-converted photons in the mode $j$ is represented by $q_{j}$.
The states in the set $\{\,\ket{l}\lsub{j}\}$ are orthonormal,
that is $\lsub{j}\braket{l}{l'}\lsub{j}=\delta_{ll'}$. We use
these states to define the logical states of the qudits. In this
sense (\ref{stgeral}) describes a composite system of two qudits.
Each qudit is represented by a state in a Hilbert space of
dimension $D$, being $D$ the number of available paths for its
transmission through the multi slits array.

It can be seen from (\ref{stgeral}) and (\ref{wlm}), that it is
possible to create different pure states of spatial qudits if one
knows how to manipulate the pump beam in order to generate
distinct transverse profiles at the plane of the multi slits
($W(\xi;z_{A})$) \cite{GLima}. Let us now assume that, before
reaching the crystal, the pump beam pass through an unbalanced
Mach-Zehnder interferometer where the transverse profile of the
laser beam is modified differently in each arm. If the difference
between the lengths of these arms is set larger than the laser
coherence length, we will obtain an incoherent superposition of
the spatial qudits states generated by each arm.

\begin{figure}[tbh]
\begin{center}
\includegraphics[width=0.37\textwidth]{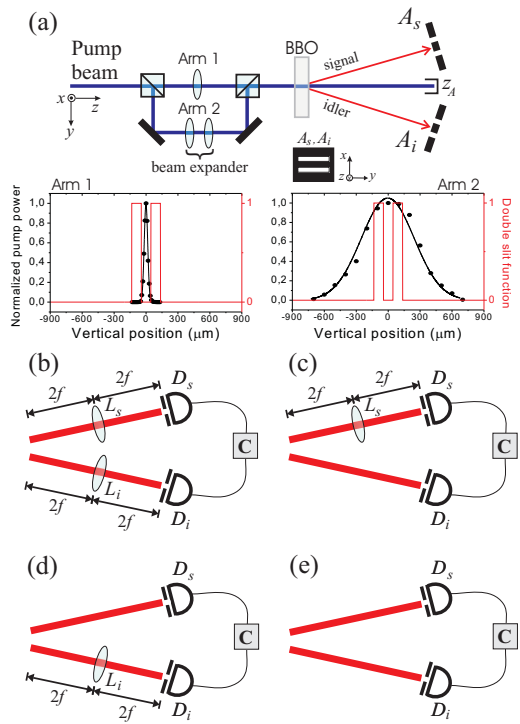}
\end{center}
\caption{(a) Schematic diagram of the experimental setup used for
generating and characterizing the mixed states of spatial qubits.
The pump beam that cross arm 1 is focused in a narrow region at
$z_{A}$ or in a broader spatial region when it cross arm 2. $A_s$
and $A_i$ are the double-slits at signal and idler propagation
paths, respectively. $D_s$ and $D_i$ are detectors and C is a
photon coincidence counter. The configuration used to determine
the diagonal elements is represent in (b). (c) and (d) were used
for the second type of measurement and (e) for the third type. All
measurements are done in the x-axes, at the detection plane.}
\label{fig:setup}
\end{figure}

We show in the following section how to use the QTR technique to
determine the density matrix of these composite systems. The state
whose the density matrix is reconstructed experimentally is a
mixed state of two spatial qubits. This state is generated with
the experimental setup represented in figure~\ref{fig:setup}(a). A
$5$~mm $\beta$-barium borate crystal is pumped by a $500$~mW
krypton laser emitting at $\lambda = 413$~nm for generating SPDC.
Before being incident at the crystal, the pump beam cross an
unbalanced Mach-Zehnder interferometer. The difference between the
lengths of each interferometer arm ($200$~mm) is set larger than
the laser coherence length ($80$~mm). Two identical double slits
$A_s$ and $A_i$ are aligned in the direction of the signal and
idler beams, respectively, at a distance of $200$~mm from the
crystal ($z_A$). The slits' width is $2a = 0.09$~mm and their
separation, $d = 0.18$~mm. The smaller dimension of the
double-slits are in the x-direction. All measurements are done in
the x-axes, at the detection plane. At the arm 1 of the
interferometer, we place a lens that focus the laser beam at the
plane of these double slits, into a region smaller than $d$. In
arm 2, we use a set of lenses that increases the transverse width
of the laser beam at $z_A$. The transverse profiles generated are
illustrated in figure~\ref{fig:setup}(a). The photons transmitted
through the double-slits are detected in coincidence between the
detectors $D_i$ and $D_s$. Two identical single slits of dimension
$5.0$ x $0.1$ mm and two interference filters centered at $826$ nm
with $8$ nm full width at half maximum (FWHM) bandwidth are placed
in front of the detectors.

By using (\ref{stgeral}) and (\ref{wlm}), we can show that the
two-photon state, after the double slits, when only arm 1 is open
is given by
\begin{equation}
\label{qubittomo}
\ket{\Psi}_1 =
\frac{1}{\sqrt{2}}(\ket{+}_s\ket{-}_i+\ket{-}_s\ket{+}_i).
\end{equation}
To simplify, we used the state $\ket{+}_j$ and $\ket{-}_j $  in
place of the states $\ket{\frac{1}{2}}_j$ and
$\ket{\frac{-1}{2}}_j $. Thus, the state $\ket{+}_j$ ($\ket{-}_j$)
represents the photon in mode $j$ being transmitted by the upper
(lower) slit of its double slit. The state of (\ref{qubittomo}) is
a maximally entangled state of two spatial qubits and their
correlation is such that when the idler photon passes through the
upper (lower) slit of its double slit, the signal photon will only
pass through its lower (upper) slit, and vice-versa.

However, if the laser beam cross only arm 2, the state of the twin
photons transmitted by these apertures will be given by
\begin{eqnarray}
\label{prodtomo}
\ket{\Psi}_2 &= &\frac{1}{2}
\,e^{i\phi}(\ket{-}_s\ket{+}_i+\ket{+}_s\ket{-}_i)
\nonumber \\
& &+ \frac{1}{2}(\ket{-}_s\ket{-}_i + \ket{+}_s\ket{+}_i),
\end{eqnarray}
where $\phi = k d^{\,2}/8 z_{A}$. And now the correlation of this
spatial qubits is different since we can also have both photons of
the pair generated, crossing their upper or lower slits,
simultaneously.

Therefore, the two-photon state generated in our experiment when
the two arms are liberated, is a mixed state of the spatial
maximally entangled state shown in (\ref{qubittomo}) and the state
of (\ref{prodtomo}). It is described by the density operator
\begin{equation}
\label{OPDENSTOMO}
\rho_{\textrm{the}} = A \ket{\Psi}\lsub{\,1\,1\!}\bra{\Psi} + B
\ket{\Psi}\lsub{\,2\,2\!}\bra{\Psi}.
\end{equation} where A and B are the probabilities for generating the
states of arm 1 and arm 2, respectively.

\section{Reconstruction}
\label{sec:reconstruction}

Now, we show how QTR can be experimentally implemented to
reconstruct the density operator of the state (\ref{OPDENSTOMO})
generated in our setup without the use of any information about
the scheme used for this generation.

Let us briefly review the process of quantum tomography. The
diagonal elements of any density operator can be measured
directly. Therefore, quantum tomographic reconstruction is a
protocol to determine the non-diagonal elements. It consists in
the use of known unitary transformations on the system. Each
transformation generates a new density operator whose diagonal
elements are a combination of the coefficients of the
transformation and of the non-diagonal elements of the original
density operator. These new diagonal elements can be measured. The
iteration of this procedure creates a set of equations which
allows the determination of the non-diagonal elements of the
initially unknown density operator. For a detailed account on this
subject we refer the reader to \cite{Blum,Leon}.

To characterize the density operator generated in our experiment,
we first adopt a general form for it

\begin{equation}
\rho =\sum_{l,m=\pm}\rho _{l_{s}l_{i}m_{s}m_{i}}|\,l_{s},l_{i}\,\rangle \left\langle
m_{s},m_{i}\right\vert .
\label{rhoger}
\end{equation}

As we are dealing with two qubits states, the total number of
measurement basis necessary for the QTR is nine \cite{sqt}. They
can be generated by using three basis for each qubit. Since we
have spin 1/2 like systems, these basis are the eigenvectors of
the Pauli operators $\{\sigma _{x},\sigma _{y},\sigma _{z}\}$. In
our case, the eigenvectors of $\sigma _{z}$ are the slit's states
given by (\ref{base}). They form the logical base, such that
$U_{j}^{(z)}=\mathbb{I}_{j}$. Measurements in basis $\{\sigma
_{x},\sigma _{y}\}$ in mode $j$ require local unitary operations
$\{U_{j}^{(x)},U_{j}^{(y)}\}\,$, which allow us for going from
$\sigma_z$ eigenvectors to ${\sigma_x,\sigma_y}$ eigenvectors,
respectively. The density operator under these local
transformations can be written as

\begin{equation}
\rho ^{(\lambda,\mu)}=U_{s}^{(\lambda)}\otimes U_{i}^{(\mu)}\rho
U_{s}^{(\lambda)\dagger }\otimes U_{\mu}^{(y)\dagger },
\end{equation}
with $\lambda,\mu=x,y$. Since the $\sigma _{z}$ eigenvectors are the slit's states, the question which remains is: How one can implement the
above discrete local operations in these slit's states to perform the QTR of $\rho$? For answering this, we first consider the state of a photon
crossing a given slit $l$ ($l=\pm $) along mode $j$ ($j=s,i$), and propagating through the free space to a detection plane located at position
$z$. This state can be calculated by the method presented in \cite{Leo} and is described by

\begin{equation}
|\,{g_{l}}\,\rangle _{j}=\sqrt{\frac{a}{\pi }}\int \!\!d{q_{j}}\,\exp(-i\alpha
q_{j}^{2})\exp(-ilq_{j}d)\mbox{\hspace{1.3pt}sinc}\,{(q_{j}a)}|\,{1q_{j}} \,\rangle.  \label{g+-}
\end{equation} It corresponds to the free evolution of the state $|l\rangle_j$ generated by the unitary operator $U_j$ (restricted to the one-photon subspace),
that is $|\,{g_{l}}\,\rangle _{j}=U_{j}|\,{l}\,\rangle _{j}$. The operator $U_j$ is given by
\begin{equation}
U_{j}=\exp(-ik(z-z_{A}))\int \!\!d{q}\,\exp(-i\alpha q^{2})|\,{1q}\,\rangle _{jj}\langle
\,{1q}\,|,
\label{Free_evolution}
\end{equation} where $\alpha =(z-z_{A})/2k$ .

Therefore, if we propagate the state $\rho$ from the plane-$z_{A}$
to plane-$z$ the result is

\begin{equation} \label{pz}
\rho _{_{Z}}=\sum_{l,m=+,-}\rho _{l_{s}l_{i}m_{s}m_{i}}|g_{l_{s}},g_{l_{i}}\,\rangle
\left\langle g_{m_{s}},g_{m_{i}}\right\vert ,
\end{equation}
and it becomes clear that $\rho $ and $\rho _{_{Z}}$ have the same coefficients, and thus, that one can reconstruct $\rho$ by determining $\rho
_{z}$, i.e, by doing the measurements in the detection plane-z.

It can be deduced from (\ref{g+-}) that photons spread out along the measurement plane, so that we have passed from discrete variables, states
$|\,{l}\,\rangle _{j}$, to a continuously distributed state $|\,{g_{l}}\,\rangle _{j}$. Thus, for carrying out the measurement in discrete basis
in plane-$z$, we need to implement an adequate postelection process, which will allow us to recover the discrete nature of the logical states.
As we shall show in the following lines, this can be properly done by allocating at the detection plane two new slits for each mode j.

In order to explicitly show how the transformations
$U_{j}^{(\lambda )}$ and $U_{j}^{(\mu )}$ ($\lambda ,\mu =x,y$)
can be implemented to do the QTR for $\rho _{z}$, we first write
an arbitrary two photon pure state in the transverse plane-$z$ as

\begin{eqnarray}
|\,{\Psi }\,\rangle _{z} &\propto & \sum_{m,n=\pm}A_{m,n}|\,{g_{m}}\,\rangle
_{s}|\,{g_{n}} \,\rangle _{i} \label{psizg+_g-}
\end{eqnarray}

The transmitted state through the double slits placed at the detection plane is

\begin{equation} \label{transmitted s}
|\,{\Psi }_{T_{\lambda \mu }}\,\rangle _{z}=\itg{q_{s}} \itgf{q_{i}} F_{T_{\lambda \mu
}}(q_{s},q_{i})|\,1_{q_{s}}\,\rangle |\,1_{q_{i}}\,\rangle,
\end{equation}
where the transmitted state biphoton amplitude is given by
\cite{Leo}

\begin{equation}
F_{T_{\lambda \mu }}(q_{s},q_{i})=\itg{q'_{s}} \itgf{q'_{i}}F(q_{s}^{\prime
},q_{i}^{\prime })T_{\lambda }(q_{s}^{\prime }-q_{s})T_{\mu }(q_{i}^{\prime }-q_{i}).
\label{F_transmitted}
\end{equation} $T(q_{j})$ is the Fourier transform of the double slits transmission function at mode $j$

\begin{equation}
T_{\mu }(q_{j})\propto \left( e^{iq_{j}x_{\mu ,0}}+e^{iq_{j}x_{\mu ,1}}\right) %
\mbox{\hspace{1.3pt}sinc}\left( q_{j}b\right),   \label{FT}
\end{equation} where $x_{\mu ,k}$ is the position of the slit $k$ ($k = 0 , 1$) in mode $j$ at the plane-$z$. By replacing the expressions
for the states $|\,{g_{l}}\,\rangle _{j}$ in $|\,{\Psi }\,\rangle _{z}$, we determine the biphoton amplitude $F(q_{s},q_{i})$ in
(\ref{psizg+_g-}). By inserting $F(q_{s},q_{i})$ and (\ref{FT}) into (\ref{F_transmitted}), we obtain $F_{T_{\lambda \mu }}(q_{s},q_{i})$. After
a straightforward derivation, we can rewrite the two photon transmitted state (\ref{transmitted s}) as

\begin{eqnarray}
|\,{\Psi }_{T_{\lambda ,\mu }}\,\rangle _{z} &\varpropto &
\sum_{k,l=0,1} B_{k,l}|f(x_{\lambda,k})\rangle _{s}|f(x_{\mu ,l})
\rangle _{i}, \label{Psi_transmitted}
\end{eqnarray}
where

\begin{eqnarray}
B_{k,l}= \sum_{m,n=\pm} r_{m}(x_{\lambda ,k})r_{n}(x_{\mu ,l}) A_{m,n}.
\end{eqnarray}
The states of the post-selected photons which crossed this
additional pair of slits are described by
\begin{widetext}
\begin{equation}
|f(x_{\mu ,k})\,\rangle _{_{\!j}}\equiv \sqrt{\frac{b}{\pi }}\int
\!\!dq_{j}^{\prime }\,\exp(-iq_{j}^{\prime }x_{\mu ,k})\mbox{%
\hspace{1.3pt}sinc}\,\left( q_{j}^{\prime }\frac{x_{\mu k}}{2\alpha }%
+q_{j}^{\prime }b\right) |1q_{j}^{\prime }\,\rangle , \label{new_slits_states}
\end{equation}
\end{widetext} where $x_{\mu ,k}$ is the position of the slit
$k$ ($k = 0 , 1$) in mode $j$ at the plane-$z$. The transversal
position of these slits determines which effective unitary
operation was performed at the transmitted photon. We have also
defined

\begin{equation}
r_{\pm }(x_{\mu ,k})=\exp\left(i\frac{\left( x_{\mu ,k}\mp d\right) ^{2}}{4\alpha }\right)%
\mbox{\hspace{1.3pt}sinc}\,\left( \frac{\left( x_{\mu ,k}\mp d\right) }{%
2\alpha }\right).
\end{equation}

Here, by comparing (\ref{Psi_transmitted}) and (\ref{psizg+_g-}),
it can be observed that the post selection process acts on the
$|\,g_{\pm }\,\rangle _{j}$ states with the following effective
transformation
\begin{equation}
U_{j}^{(\mu)} |\,{\tilde{g}_{\pm }}\,\rangle _{j} \varpropto r_{\pm }(x_{\mu
,0})|f(x_{\mu ,0})\rangle _{j} + r_{\pm }(x_{\mu ,1})|f(x_{\mu ,1})\rangle _{j},
\label{U_effective}
\end{equation}
where $|\,{\tilde{g}_{\pm }}\,\rangle _{j}{\ }$ state denotes the
post selected state arising from $|\,{g_{\pm }}\,\rangle _{j}$
state. By considering the value of the experimental parameters:
$z-z_{A}$, $d$, $a$ and $b$, it can be shown that the states
$|f(x_{\mu ,0})\rangle _{j}$ and $|f(x_{\mu ,1})\rangle _{j}\ $
are orthogonal when the condition $\left\vert x_{\mu ,1}-x_{\mu
,0}\right\vert>4b$ is satisfied.

The positions of the new slits for generating the effective
transformation $U_{j}^{(x)}$ ($U_{j}^{(y)}$) are $x_{x,0}=0$
($x_{y,0}=-\Delta /2$) and $x_{x,1}=\Delta $ ($x_{y,1}=\Delta
/2$), with $\Delta =\frac{\alpha \pi }{d}=1376$ mm (Note that
condition $\left\vert x_{\mu ,1}-x_{\mu ,0}\right\vert =\Delta
>4b$ is widely satisfied). We remark that we refer to the
transformations done by these slits as effective transformations,
due to the fact that they act as unitary transformations only for
the photons in mode $j$ which were transmitted through the pair of
slits at the detection plane.

Because of the linearity of quantum mechanics and because we are
performing only local operations to the twin photons, we know that
the diagonal elements of $\rho ^{(\lambda ,\mu )}$ operators are
linear combinations of the coefficients $\rho
_{l_{s}l_{i}m_{s}m_{i}}$ of (\ref{pz}). The diagonal elements of
the transformed density operators $\rho ^{(\lambda ,\mu )}$ at the
plane-z are simply determined by using four coincidence numbers
measured when $D_{s}$ is at the position $x_{\lambda ,k}$ for
$k=0,1$ and $D_{i}$ is fixed at the transversal position $x_{\mu
,0}$\ or at $x_{\mu ,1}$. We assume that detectors $D_{s}$ and
$D_{i}$ are placed just behind the new slits at the plane-z. These
diagonal elements are then
\begin{equation}
\rho _{L_{s}L_{i},L_{s}L_{i}}^{(\lambda ,\mu )}=\sum_{l_{s}l_{i}m_{s}m_{i}= \pm
}U_{L_{s},l_{s}}^{(\lambda )}U_{L_{i},l_{i}}^{(\mu )}\rho
_{l_{s}l_{i}m_{s}m_{i}}U_{m_{s},L_{s}}^{(\lambda )}U_{m_{i},L_{i}}^{(\mu )},
\label{new_diagonal}
\end{equation}
where $L_{s},L_{i}=0,1$ for $\lambda,\mu=x,y$ or $L_{s},L_{i}=\pm$
for $\lambda,\mu=z$, and $U_{L_{j},l_{j}}^{(\mu)}$ are
coefficients of the effective transformation $U_{j}^{(\mu)}$ given
by (\ref{U_effective}). One can now obtain the non-diagonal
elements of the density operator, $\rho$, just by inverting the
above linear equations. Besides, in case of a QTR for spatial
qudits, the number of necessary slits at the plane-$z$ is equal to
the dimension $D$ of the qudits. The positions of these slits can
be determined by using the eigenvectors of the $D^{\,2}-1$
generators of $su(D)$ algebra \cite{algebra}.

\subsection{Diagonal Elements}

The measurement in the logical base, namely $\rho ^{(z,z)}$, can
be determined by coincidence measurements with the detectors just
behind the double slits \cite{GLima} or at the plane of image
formation when lenses are placed in the path of the double-slits
transmitted photons as showed in figure~\ref{fig:setup}(b)
\cite{GLima2}. In this last configuration, when the detector
$D_{j}$ is at position $x_{z,0}=-100$ $\mu $m ($x_{z,1}=+100$ $\mu
$m), it detects all photons that cross the slit $+$ ($-$). By
doing these measurements and normalizing the coincidences recorded
for the four slits we obtained: $\rho _{++,++}=0.028$, $\rho
_{+-,+-}=0.468$, $\rho_{-+,-+}=0.462$ and $\rho _{--,--}=0.042$.

\subsection{Non diagonal Elements}

The second type of coincidence measurements were done by
positioning the signal detector at $x_{z,0}=-100$ $\mu $m  or
$x_{z,1}=+100$ $\mu $m and with the idler detector in the plane-z
at the transversal positions $x_{y,0}=-\Delta /2$ ($-0.688$mm),
$x_{y,1}=\Delta /2$ ($0.688$mm), $x_{x,0}=0\,$mm and
$x_{x,1}=\Delta $ ($1.376\,$mm) (See figure~\ref{fig:setup}(c)).
When the idler detector is at the transverse position $x_{y,0}$ or
$x_{y,1}$, the detector selects the idler photons in the
$|f(x_{y,0})\rangle _{i}$ state or in the$|f(x_{y,1})\rangle _{s}$
state. When the idler detector is at the transverse position
$x_{x,0}$ or $x_{x,1}$, it detects the idler photons in the
$|f(x_{x,0})\rangle _{i}$ state or in the $|f(x_{x,1})\rangle
_{s}$ state. With these eight measured coincidence numbers, we
determined the non diagonal elements of the density operator
$\{\rho _{++,+-},\rho _{+-,++},\rho _{-+,--},\rho _{--,-+}\}$. By
repeating this detection procedure and reversing the roles of the
signal and idler detectors (See figure~\ref{fig:setup}(d)), we
found the non diagonal elements $\{\rho _{++,-+},\rho
_{-+,++},\rho _{+-,--},\rho _{--,+-}\}$. We show below the
explicit expressions that determine $\rho _{++-+}$

\begin{eqnarray}
\re{\rho _{++-+}} &=&\frac{\rho _{0+,0+}^{(x,z)}-\rho _{++++}\cos ^{2}\theta
_{x}-\rho _{-+-+}\sin ^{2}\theta _{x}}{\sin 2\theta _{x}},  \label{real} \nonumber \\
\end{eqnarray} and

\begin{eqnarray}
\im{\rho _{++-+}} &=& \frac{-\rho _{0+,0+}^{(y,z)}+\rho _{++++}\cos ^{2}\theta _{y} +
\rho _{-+-+}\sin ^{2}\theta _{y}}{\sin 2\theta _{y}}
\label{imag}, \nonumber \\
\end{eqnarray} where

\begin{equation}
\cos \theta _{\mu }=\frac{\left\vert r_{+}(x_{\mu ,0})\right\vert }{\sqrt{%
\left\vert r_{+}(x_{\mu ,0})\right\vert ^{2}+\left\vert r_{-}(x_{\mu ,1})\right\vert
^{2}}}.
\end{equation}
The above expressions for $\re{\rho _{++-+}}$ and $\im{\rho
_{++-+}}$ are obtained by inverting Eq. (\ref{new_diagonal}), with
$L_s=0,L_i=+$, both for ${\lambda=x,\mu=z}$ and for
${\lambda=y,\mu=z}$.

In the third measurement type shown in figure~\ref{fig:setup}(e),
signal and idler detectors are positioned in the  detection
plane-$z$ at the positions $x_{\lambda ,k}$ and $x_{\mu ,l}$, with
$\lambda ,\mu $ being $x$ or $y$ and $k,l$ being $0$ or $1$. This
allows, by means of similar expressions to (\ref{real}) and
(\ref{imag}), the determination of $\{\rho _{++,--},\rho
_{--,++},\rho _{+-,-+},\rho _{-+,+-}\}$. This set of measurements
correspond to local operations being applied to each of the
down-converted photons, simultaneously.

\subsection{The Reconstructed Density Operator}

By performing the quantum tomographic reconstruction, as described
above, we found the following form for the density operator in its
matrix representation of our experiment
\begin{widetext}
\begin{equation}
\qquad \rho = \left[\begin{array}{cccc}
0.028 &0.083 + 0.004i &0.081 + 0.005i &-0.129 + 0.062i \\
0.083 - 0.004i &0.468 &0.444 - 0.058i &0.097 - 0.008i \\
0.081 - 0.005i &0.444 + 0.058i &0.462 &0.096 - 0.006i \\
-0.129 - 0.062i &0.097 + 0.008i &0.096 + 0.006i &0.042 \\
\end{array}\right]. \label{rhoexp}
\end{equation}
\end{widetext} The elements of a density operator must satisfy
the Schwarz inequality, i.e., $|\rho_{jk}|\leq\sqrt{\rho_{jj}
\rho_{kk}}$, where $j,k=++$, $+-$, $-+$ and $--$, if it really
represents a quantum state. This is not our case for the matrix
element $\rho_{++--}$, since it can be seen that $|\rho_{++--}|>
\sqrt{\rho_{++++} \rho_{----}}$. The reason for that are the
experimental fluctuations present in the coincidence measurements
which can affect the final result as we discussed in the
Introduction. This discrepancy can be reduced by increasing the
detection time. Even though our reconstructed density matrix
presents properties which are not fully compatible with the
quantum state description, it is possible to show that it is
consistent with the theory developed in
section~\ref{sec:generation}. This is done in the next section,
where we also show experimental evidences of the good quality of
our reconstruction.

\section{Discussion and Conclusion}
\label{sec:discussion}

The measured density operator shown in (\ref{rhoexp}) can be
approximately written as
\begin{equation}  \label{OPDENSTOMOExp}
\rho = 0.87 |\,{\Phi}\,\rangle _{_{\!\scriptstyle \,1\,1\!}} \langle\,{\Phi}%
\,| + 0.13 |\,{\Phi}\,\rangle _{_{\!\scriptstyle \,2\,2\!}} \langle\,{\Phi}%
\,| ,
\end{equation}
where the states $|\,{\Phi}\,\rangle $ are given by

\begin{eqnarray}  \label{phimes}
|\,{\Phi}\,\rangle _1 &= &0.077 e^{i\phi_1} |\,{++}\,\rangle + 0.704
e^{i\phi_2}|\,{+-}\,\rangle  \nonumber \\
& &+ 0.699 e^{i\phi_2}|\,{-+}\,\rangle + 0.099 e^{i\phi_3} |\,{--}\,\rangle ,
\end{eqnarray}
and

\begin{eqnarray}  \label{phiparc}
|\,{\Phi}\,\rangle _2 &= &0.514 |\,{++}\,\rangle + 0.502 e^{i\theta}|\,{+-}%
\,\rangle  \nonumber \\
& &+ 0.501 e^{i\theta}|\,{-+}\,\rangle + 0.483|\,{--}\,\rangle ,
\end{eqnarray}
with $\phi_1 \simeq \phi_2 \simeq \phi_3 \approx 4.2$ and
$\theta=0.07$.

However, the possibility to decompose the density operator,
$\rho$, in terms of the projectors of a state, $|\,{\Phi}\,\rangle
_1$, which has a high degree of entanglement and a state,
$|\,{\Phi}\,\rangle _2$, that is of the form predicted by
(\ref{prodtomo}), is not sufficient for associating them with the
states generated by each arm of the interferometer in our
experiment. We still have to give an experimental evidence which
corroborates with (\ref{OPDENSTOMOExp}) as a reasonable
approximation for the quantum state of the twin photons, i.e., we
need to show that the values of $A=0.87$ and $B=0.13$, obtained
mathematically, are reasonable for the probabilities of generating
these states in each arm.

We measured the values of $A$ and $B$ by blocking one of the arms
of the interferometer and detecting the transmitted coincident
photons through the signal and idler double-slits. A (B) is the
ratio between the coincidence rate when arm 2 (arm 1) is blocked
and the total coincidence rate when both arms are unblocked. From
this measurement we obtained, $A = 0.85 \pm 0.03$ and $B = 0.15
\pm 0.03$.

\begin{figure}[t]
\begin{center}
\includegraphics[width=0.27\textwidth]{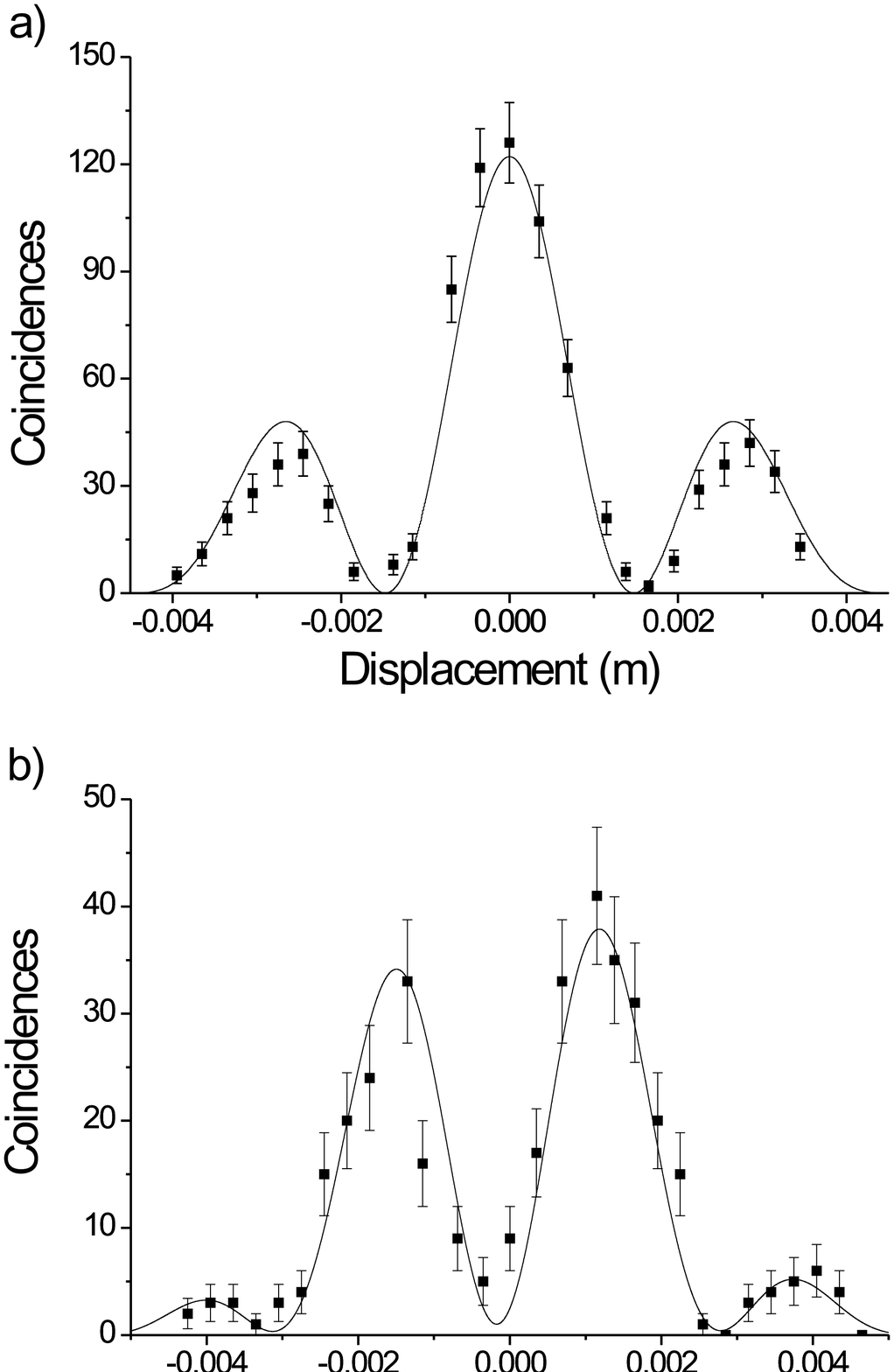}
\end{center}
\caption{Fourth order interference pattern as a function of
$D_{s}$ position. In (a), the detector idler was fixed at the
transverse position $x = 0$~mm. In (b), it was fixed at the
transverse position $x = 1376$~mm. The solid curves were obtained
theoretically.} \label{fig:Conditional}
\end{figure}

Another experimental evidence for the high value of $A$ can be
found in the fourth order interference pattern recorded (See
figure~\ref{fig:Conditional}). The interference pattern is
recorded by using the configuration shown in
figure~\ref{fig:setup}(e). Fourth order interference pattern as a
function of $D_{s}$ position was recorded. In (a), the detector
idler was fixed at the transverse position $x = 0$~mm. In (b), it
was fixed at the transverse position $x = 1376$~mm. The solid
curves were obtained theoretically \cite{Fonseca}, with $A$ and
$B$ as free parameters. Since the state $|\,{\Phi}\,\rangle _1$ is
almost a maximally entangled state we would expect to observe
conditional interference patterns \cite{Fonseca,Greenberger} when
both interferometer arms are unblocked. This would not be the case
for high values of $B$. The conditionality can be clearly observed
in our interference patterns. The reason for having the
probability of generating the state from arm 1 much higher than
the probability of generating the state of arm 2 is quite simple.
The laser beam that cross arm 1 of the interferometer is focused
at the double slit's plane-$z_A$, and the spatial correlation of
the generated photons is such that it is more favorable to the
transmission of the twin photon through the slits than it is when
the photon pairs are generated by the pump beam that cross arm 2
\cite{Leonardo}. These values can be properly manipulated by
inserting attenuators at the interferometer.

\begin{figure}[t]
\begin{center}
\includegraphics[width=0.45\textwidth]{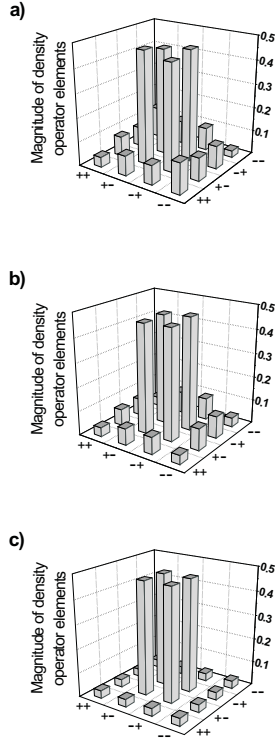}
\end{center}
\caption{Histogram of the real part of the matrix elements for (a)
the measured density operator, (b) the density operator given by
(\protect \ref{OPDENSTOMOExp}) and (c) the predicted density
operator of Sec.~ \protect \ref{sec:generation}.}
\label{fig:histograma}
\end{figure}

These experimental observations confirm the good quality of the
QTR performed on the two photon state and allow us to consider the
states $|\,{\Phi}\,\rangle _1$ and $|\,{\Phi}\,\rangle _2$ as good
approximations for the states generated by arm 1 and arm 2 of the
interferometer used. Figure~\ref{fig:histograma} shows a histogram
of the real part of the matrix elements of (a) the measured
density operator of (\ref{rhoexp}), (b) the density operator given
by (\ref{OPDENSTOMOExp}) and, (c) the predicted density operator
of section~\ref{sec:generation}. The agreement between the
predicted and the measured density operator is good within the
experimental errors. The largest error for the diagonal elements
is only $3.5\%$. But, for the non-diagonal elements the propagated
errors reaches $30\%$ for their real parts and up to $65\%$ for
the imaginary parts. Again we remember that these errors can be
decreased by increasing the detection time.

In conclusion, we have demonstrated that it is possible to
generate a broad family of mixed states of spatial qudits by
exploring the transverse correlation of the down-converted
photons. A statistical mixture of spatial qubits were used to show
the quantum tomographic reconstruction performed to measure its
density operator. The process was discussed in details and
experimental evidences for the good quality of the reconstruction
performed were showed. Even though we had considered the state
determination only for the case of qubits, it can be generalized
and performed in a similar way for higher dimension systems in a
mixed state. The importance of this work comes from the
possibility of using spatial qudits for quantum communications
protocols, where it requires the ability to characterize them in
the presence both of noise source and of an undesired user at the
quantum communication channel.

\section*{ACKNOWLEDGMENT}

The authors would like to express their gratitude to Marcelo T.
Cunha for having called their attention to this problem and
initiating the discussions which culminated in this work. G. Lima,
L. Neves and S. P\'adua were supported by CAPES, CNPq, FAPEMIG and
Instituto do Mil\^enio de Informa\c{c}\~ao Qu\^antica. C. Saavedra
and A. Delgado were supported by Grants Nos. FONDECYT 1061046 and
Milenio ICM P02-49F. F. Torres was supported by MECESUP UCO0209.
This work is part of the international cooperation agreement
CNPq-CONICYT 491097/2005-0.

\end{document}